\documentclass[final,5p,times,twocolumn,authoryear]{elsarticle}

\usepackage{amssymb}
\usepackage{amsmath}
\usepackage{array}
\usepackage{float}
\usepackage{graphicx}
\usepackage{lipsum}
\usepackage{xcolor}
\usepackage{makecell}
\usepackage{multirow}
\usepackage{tabularx}
\usepackage{url}

\defcitealias{Guidorzi15a}{G15}

\journal{Astronomy $\&$ Computing}

\begin{document}

\begin{frontmatter}

\title{{\sc fast-mepsa}: an optimised and faster version of peak detection algorithm {\sc mepsa}}

\author[first,second]{Manuele Maistrello}
\author[first,second]{Romain Maccary}
\author[first,second,third]{Cristiano Guidorzi}
\affiliation[first]{organization={University of Ferrara, Department of Physics and Earth Sciences},
            addressline={via Saragat 1}, 
            city={Ferrara},
            postcode={44122}, 
            state={},
            country={Italy}}

\affiliation[second]{organization={Istituto Nazionale di Astrofisica, Osservatorio di Astrofisica e Scienza dello Spazio di Bologna},
            addressline={via Gobetti 93/3}, 
            city={Bologna},
            postcode={40129}, 
            state={},
            country={Italy}}

\affiliation[third]{organization={Istituto Nazionale di Fisica Nucleare, Sezione di Ferrara},
            addressline={via Saragat 1}, 
            city={Ferrara},
            postcode={44122}, 
            state={},
            country={Italy}}

\begin{abstract} % 250 parole

We present {\sc fast-mepsa}, an optimised version of the {\sc mepsa} algorithm developed to detect peaks in uniformly sampled time series affected by uncorrelated Gaussian noise. Although originally conceived for the analysis of gamma-ray burst (GRB) light curves (LCs), {\sc mepsa} can be readily applied to other transient phenomena. The algorithm scans the input data by applying a set of 39 predefined patterns across multiple timescales. While robust and effective, its computational cost becomes significant at large re-binning factors.
To address this, {\sc fast-mepsa} introduces a sparser offset-scanning strategy. In parallel, building on {\sc mepsa}’s flexibility, we introduce a 40th pattern specifically designed to recover a class of elusive peaks that are typically sub-threshold and lie on the rising edge of broader structures---often missed by the original pattern set.
Both versions of {\sc fast-mepsa}---with 39 and 40 patterns---were validated on simulated GRB LCs. Compared to {\sc mepsa}, the new implementation achieves a speed-up of nearly a factor 400 at high re-binning factors, with only a minor ($\sim 4\%$) reduction in the number of detected peaks. It retains the same detection efficiency while significantly lowering the false positive rate of low significance. The inclusion of the new pattern increases the recovery of previously undetected and sub-threshold peaks.
These improvements make {\sc fast-mepsa} an effective tool for large-scale analyses where a robust trade-off between speed, efficiency, and reliability is essential. The adoption of 40 patterns instead of the classical 39 is advisable when an enhanced efficiency in detecting faint events is desired. The code is made publicly available.

\end{abstract}


\end{frontmatter}

\section{Introduction}
\label{sec:introduction}

In the current time-domain era, monitoring how transient astrophysical phenomena vary over time offers key insights into the physical processes driving their emission. These temporal variations span a broad range of timescales, from milliseconds to years or even longer. Among the most energetic and enigmatic stellar-scale explosions, gamma-ray bursts (GRBs)---brief, intense flashes of gamma rays---stand out as exceptional events.

GRB light curves (LCs, i.e. their flux time profiles) encode a wealth of information about the activity and, possibly, the nature of their central engine. Most LCs appear as a sequence of pulses, often grouped into emission episodes and separated by so-called quiescent times. These pulses vary in shape, duration, and intensity, and exhibit no obvious periodicity \citep{Guidorzi25}. Several models have been proposed to explain this diversity(see \citealt{KumarZhang15rev} for a review). In particular, whether LCs arise from a common stochastic process remains an open question. A precise characterisation of any such underlying process therefore requires robust peak‐detection methods. This underscores the importance of a method capable of reliably identifying as many peaks as possible in observed LCs, while properly accounting for statistical uncertainties.

In this context, \citet[hereafter G15]{Guidorzi15a} introduced a well-calibrated peak search algorithm named {\sc mepsa} (Multiple Excess Peak Search Algorithm), which has since been widely used to characterise various LC properties---including the minimum variability timescales \citep{Camisasca23}; the search for periodic activity \citep{Guidorzi25}; the distribution of the number of peaks per GRB \citep{Guidorzi24, Maccary24b}; the energy of individual pulses \citep{Maccary24}; and the waiting time distribution \citep{Guidorzi15b}.

The simultaneous detection of GW\,170817 in gravitational waves \citep{Abbott17} and GRB\,170817A in $\gamma$ rays \citep{Goldstein17,Savchenko17} confirmed the long-suspected association between short GRBs and neutron star binary mergers. In this context, \citet{Kocevski18} and, more recently, \citet{Fletcher24} jointly analysed \textit{Fermi}/GBM and \textit{Swift}/BAT data in search of sub-threshold GRBs---i.e. events that not initiate an on-board trigger---temporally coincident with LIGO/Virgo compact binary coalescence triggers. These studies leveraged external information from independent detections, highlighting the importance of robust and sensitive algorithms capable of identifying such elusive events.

Motivated by these needs, we present {\sc fast-mepsa}, an optimised version of the {\sc mepsa} algorithm that preserves its operative structure and philosophy. {\sc fast-mepsa} has been specifically designed to achieve an optimal balance between computational speed and detection sensitivity, making it particularly suited to large-scale searches. In parallel, we propose a new pattern that, when added to the 39 already present, contributes to enhance the capability of the code to identify faint and sub-threshold events, that would be mostly missed otherwise.
 
The main features of the original method are summarised in Section~\ref{subsec:mepsa}, while the modifications introduced in {\sc fast-mepsa} are described in Section~\ref{subsec:mepsa_fast}. In Section~\ref{sec:test_and_validation}, we validate the new algorithm and compare its performance with that of the original {\sc mepsa}. Conclusions are presented in Section~\ref{sec:conclusions}. The C code and pattern sets are publicly available\footnote{\url{https://www.fe.infn.it/u/guidorzi/new_guidorzi_files/code.html}}.

\section{Algorithm description}
\label{sec:algorithm_description}

\subsection{The original algorithm}
\label{subsec:mepsa}

{\sc mepsa} is designed to identify peaks in a uniformly sampled time series affected by uncorrelated Gaussian noise. Specifically, it applies a set of 39 patterns simultaneously to each bin $i$ of the input LC. Each pattern $P_k$ comprises $n_{k,l}$ leftward and $n_{k,r}$ rightward bins relative to bin $i$. Each of these surrounding bins is assigned a threshold $v_{k,j}$, with $j = 1, \ldots, n_{k,l} + n_{k,r}$. A pattern $P_k$ is considered fulfilled at bin $i$ if
\begin{equation}
    \begin{cases}
        r_i - r_j \geq v_{k,(j - i + n_{k,l} + 1)} \, \sigma_{i,j}, & j = i - n_{k,l}, \ldots, i - 1, \\
        r_i - r_j \geq v_{k,(j - i + n_{k,l})} \, \sigma_{i,j}, & j = i + 1, \ldots, i + n_{k,r},
    \end{cases}
\end{equation}
where $r_i$ is the rate in bin $i$, and $\sigma_{i,j} = (\sigma_i^2 + \sigma_j^2)^{1/2}$, with $\sigma_i$ denoting the statistical noise in bin $i$. Bin $i$ is promoted to a peak candidate if at least one pattern is satisfied. 

The same screening criteria are then applied to re-binned versions of the input LC, with the re-binning factor $F_{\rm reb}$ increasing linearly up to a maximum value $F_{\rm reb,m}$. Furthermore, for each $F_{\rm reb}$, the algorithm scans all possible offsets of the LC: there are exactly $F_{\rm reb}$ distinct starting bins and as many different re-binned profiles. Finally, whenever the same peak candidate is detected multiple times, {\sc mepsa} performs a cross-check by comparing the detection times and associated timescales, retaining the one with the highest statistical significance.

\subsection{The faster algorithm}
\label{subsec:mepsa_fast}

\citetalias{Guidorzi15a} showed that the computational time of {\sc mepsa} scales approximately as $F_{\rm reb,m}^3$ when applied to typical GRB LCs with average SNR and binning times of a few milliseconds. This scaling imposes a practical constrain the maximum re-binning factor that can be adopted. 
A detailed analysis of their findings reveals that approximately 64\% of the peaks detected by {\sc mepsa} are associated with a re-binning factor less than or equal to 10, while only 0.5\% correspond to factors greater than 100. 

Guided by these findings, we introduced two key modifications to reduce computational cost. First, we modified the way the re-binning factor increases:
\begin{equation}
    F_{\rm reb}(n) = 
    \begin{cases}
        n, & 1 \leq n \leq 10, \\
        a n^2 + b n + c, & 10 < n \leq R,
    \end{cases}
\end{equation}
where $n$ is an integer variable and 
\begin{equation}
    R = \left\lfloor \dfrac{-b + \sqrt{b^2 - 4a(c - F_{\rm reb,m})}}{2a} \right\rfloor,
\end{equation}
with $b = 1 - 20a$ and $c = 100a$. These constraints ensure continuity at the transition between the linear and parabolic regimes. As continuity provides only two conditions, one degree of freedom remains for defining the parabolic growth; we set $a = 1/2$, which yields:
\begin{equation}
    F_{\rm reb}(n) =
    \begin{cases}
        n, & 1 \leq n \leq 10, \\
        \dfrac{1}{2} n^2 - 9n + 50, & 10 < n \leq R.
    \end{cases}
\end{equation}
Choosing $a = 1/2$ ensured that the scanning becomes sparser at higher $F_{\rm reb}$, but not excessively. This choice reduces the computational time without compromising detection efficiency, as discussed in Section~\ref{sec:test_and_validation}.

Secondly, we modified the set of LC offsets scanned by the algorithm. Let $K$ be the smallest integer such that $K\ge F_{\rm reb} / 10$. The sequence of scanned offsets $O_i$ is then defined recursively as:
\begin{equation}
    \begin{cases}
        O_0 = 0, \\
        O_{i+1} = O_i + K,
    \end{cases}
\end{equation}
with $i = 0, \ldots, \left\lfloor F_{\rm reb} / K \right\rfloor$, and $K$ being the step size of the series. With this scheme, the algorithm scans all possible LC offsets up to a re-binning factor of 10, while the scanning becomes progressively sparser for greater $F_{\rm reb}$. 

Hereafter, we refer to this faster version of the code as {\sc fast-mepsa}.

\subsubsection{The 40th pattern}
\label{subsec:40_pattern}

\citet{Maccary24} showed that around 10\% of the pulses in their sample that were visually identified, were missed by {\sc mepsa}. We carried out a detailed characterisation of this small fraction of elusive peaks to understand whether they share any common traits. Interestingly, most of them occur on the rising part of other longer-lasting structures, and tend to be predominantly sub-threshold, low-SNR pulses. To improve the detection efficiency in such cases, we developed a new pattern specifically designed to model the rising phase of these peaks and maximise their detection rate.

With reference to Table~1 in \citetalias{Guidorzi15a}, the corresponding thresholds are reported in Table~\ref{tab:40_pattern}. We tested both {\sc mepsa} and {\sc fast-mepsa} including the new pattern. The results are reported in Section~\ref{sec:test_and_validation}.

\subsection{Data handling}
\label{subsec:data_handling}
The following points are valid for both {\sc mepsa} and {\sc fast-mepsa}, and discuss the conditions that must be fulfilled by the input data for a correct usage. 

\subsubsection{Background subtraction}
{\sc mepsa} is designed to operate on background-subtracted time profiles. Although this procedure is not directly part of the algorithm, it has to be properly carried out prior to its application. In fact, an inaccurate background estimate can affect the performance: for example, an under-fitting can lead to false positives while an over-fitting to a sensitivity loss. In these cases, a combination of statistical tests, such as $\chi^2$ and runs tests on the residuals, can identify trends due to poor background modelling. In general, a poor knowledge and/or modelling of background inevitably affects the sensitivity and reliability of any peak detection algorithm.

\subsubsection{Poisson noise}
{\sc mepsa} relies on the assumption that the time series is dominated by Gaussian noise. This assumption holds true if the number of counts in each bin is sufficiently high for the Poisson distribution to approach the Gaussian limit. In the low-counting regime, this approximation breaks down, potentially leading to poor performances. In such cases, LCs should be re-binned to a sufficient timescale that ensures the Gaussian limit prior to the algorithm application, provided this does not affect the genuine variability of the signal. When this condition cannot be satisfied, {\sc mepsa} can still be adopted if the average counts per bin is a few, provided that a higher value on the SNR threshold is adopted, a practical approach that was occasionally adopted \citep{Guidorzi25}. Instead, when the regime is highly Poisson and non-Gaussian (that is, less than a few counts per bin), {\sc mepsa} should not be applied and other algorithms should be adopted, such as those devised in \citet{Guidorzi20a}.

\subsection{Re-binning factor}
The goal of the algorithm is to detect peaks characterised by a wide range of SNR and/or timescales. A common assumption adopted in our previous works, in which we made an extensive use of {\sc mepsa}, was to set $F_{\rm reb,m} = 256$ for \textit{CGRO}/BATSE and \textit{Swift}/BAT, and $F_{\rm reb,m} = 128$ for \textit{Fermi}/GBM. Considering a nominal bin size of 64~ms, these choices correspond to scanning the LCs up a bin-time of 16.384~s and 8.192~s, respectively. A relatively large re-binning factor is crucial to identify peaks that last a correspondingly long time and, as such, do not exceed the SNR threshold if integrated over shorter time intervals.

\begin{table*}
\setlength{\tabcolsep}{12pt}
\renewcommand{\arraystretch}{1.4}
    \centering
    \begin{tabular}{c|cccccccccccc}\hline
        $k$ & $v_{k,l}$ & $v_{k,r}$ & $v_{k,1}$ & $v_{k,2}$ & $v_{k,3}$ & $v_{k,4}$ & $v_{k,5}$ & $v_{k,6}$ & $v_{k,7}$ & $v_{k,8}$ & $v_{k,9}$ & $v_{k,10}$ \\\hline
        40 & 6 & 2 & 3.5 & 3.0 & 2.5 & 2.0 & 1.5 & 0.5 & 0.5 & 1.5 &----&----\\\hline
    \end{tabular}
    \caption{The 40th pattern. $v_{k,l}$ and $v_{k,r}$ indicate the number of bins to the left and right of the $i$th bin in the input time series, respectively.}
    \label{tab:40_pattern}
\end{table*}

\section{Test and validation}
\label{sec:test_and_validation}

We have two alternative codes, {\sc mepsa} and {\sc fast-mepsa}, along with two possible sets of patterns, either 39 or 40. Consequently, the user may choose between four different combinations of codes and patterns. Hereafter, we systematically compare and characterise each of these four possible usages.

Both algorithms were tested on the same sample of simulated LCs described in \citetalias{Guidorzi15a}. We below summarise the main properties of these time profiles:

\begin{itemize}
    \item Group 1: $N_1 = 300$ LCs, each consisting of 5000 bins of 64~ms. The counts in each bin were generated according to a Poisson distribution with expected value $1000~\mathrm{cnts/bin}$. In this way, the count rate is approximately normally distributed. These profiles are meant to represent background counts from scintillators operating approximately in the 10--1000~keV band, such as \textit{CGRO}/BATSE, \textit{BeppoSAX}GRBM, \textit{Fermi}/GBM, \textit{Insight-HXMT}/HE, and \textit{SVOM}/GRM.
    
    \item Group 2: $N_2 = 100$ LCs, each with 15000 bins of 64~ms. The counts in bin $i$ were drawn from a normal distribution\footnote{As a coded-aperture telescope, BAT produces background-subtracted GRB LCs directly \citep{Barthelmy05}. Moreover, being the result of linear combinations of the counts of many independent detectors, the rates are normally distributed.} $N(0, \sigma_i)$, with $\sigma_i$ values taken from a typical mask-weighted \textit{Swift}/BAT GRB LC in the 15--150~keV band, extracted following the standard procedure recommended by the BAT team\footnote{\url{https://swift.gsfc.nasa.gov/analysis/threads/bat_threads.html}}. As a coded-mask instrument, this case is also representative of \textit{SVOM}/ECLAIR.
    
    \item Group 3: $N_3 = 150$ LCs, each with 15000 bins of 64~ms, populated with pulses of fast-rise exponential-decay (FRED; \citealt{Norris96}) shape. Following \citetalias{Guidorzi15a}, we assumed fixed values for the peakedness $\nu = 1.5$, the rise time $\sigma_r = 1~\mathrm{s}$, and the decay time $\sigma_d = 3~\mathrm{s}$. Based on this assumption, the corresponding full width at half maximum (FWHM) equals $(\ln2)^{1/\nu}(\sigma_r + \sigma_d)$. Pulses within the same LC were generated assuming exponentially distributed waiting times, corresponding to a memoryless Poisson process with constant expected pulse rate per unit time. Specifically, we used a pulse rate ranging from $1/40$ up to $1/20$ pulses/s. Pulse amplitudes were selected so that the pulse signal-to-noise ratio (SNR) spans the range $0.5 \lesssim \log{(\mathrm{SNR})} \lesssim 2$. Background was simulated using the same templates as in Group 2.
\end{itemize}

\subsection{Computational time}
\label{subsec:runtime}

The computational time was estimated using Group 3 LCs. Each algorithm was applied to every LC, and the runtime was recorded. To assess the effectiveness of the modifications introduced in Section~\ref{subsec:mepsa_fast} at high re-binning factors, we set $F_{\rm reb,m} = 512$. The mean runtimes and their associated $1\sigma$ uncertainties are reported in Table~\ref{tab:runtime}, which highlights a reduction in execution time by a factor of approximately 400, regardless of the number of patterns used.

\begin{table}[H]
\setlength{\tabcolsep}{8pt}
\renewcommand{\arraystretch}{1.4}
    \centering
    \begin{tabular}{c|cc}\hline
         Number of patterns & {\sc mepsa} & {\sc fast-mepsa} \\\hline
         39 & $(412 \pm 13)\,\mathrm{s}$ & $(0.99 \pm 0.07)\,\mathrm{s}$ \\
         40 & $(408 \pm 6)\,\mathrm{s}$ & $(1.02 \pm 0.07)\,\mathrm{s}$ \\\hline
    \end{tabular}
    \caption{Mean computational time (with $1\sigma$ uncertainties) for each version of the algorithm, measured on Group 3 LCs.}
    \label{tab:runtime}
\end{table}

\subsection{False positive rate}
\label{subsec:fp}

As in \citetalias{Guidorzi15a}, we computed the false positive (FP) rate for each algorithm using Group 1 and Group 2 LCs, as these are purely affected by statistical noise and contain no real structures. The results, summarised in Table~\ref{tab:fp}, show that {\sc fast-mepsa} significantly reduces the number of FPs, primarily due to the sparser scanning of the input LC at large re-binning factors. More specifically, the number of FPs drops from 56 to 13 (a 77\% reduction) when using 39 patterns, and from 75 to 37 (a 50\% reduction) when using 40 patterns.

\begin{table}[H]
\setlength{\tabcolsep}{8pt}
\renewcommand{\arraystretch}{1.4}
    \centering
    \begin{tabular}{c|cc}\hline
         Number of patterns & {\sc mepsa} & {\sc fast-mepsa} \\\hline
         39 & 56 ($3.7\times10^{-5}$) & 13 ($8.7\times10^{-6}$) \\
         40 & 75 ($5.0\times10^{-5}$) & 37 ($2.5\times10^{-5}$) \\\hline
    \end{tabular}
    \caption{Number of FPs detected by each version of the algorithm in Group 1 and 2 LCs. The corresponding fraction out of $1.5\times10^6$ scanned bins is reported in brackets.}
    \label{tab:fp}
\end{table}

\begin{figure}[H]
    \centering
    \includegraphics[width=0.4\textwidth]{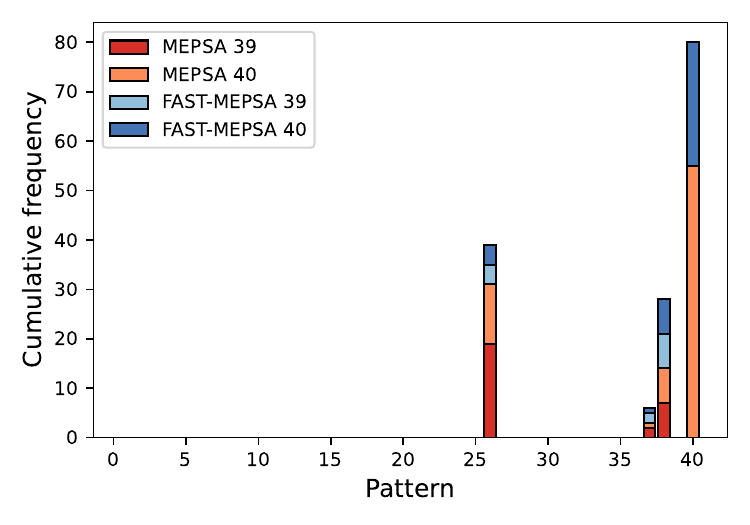}
    \caption{Cumulative histogram of the FPs detected by all four combinations of codes and patterns, as a function of the triggering pattern. Different colours represent the contribution of each algorithm to the total frequency in each histogram bin.}
    \label{fig:fp_distribution}
\end{figure}

We also analysed the distribution of FPs among the patterns. Figure~\ref{fig:fp_distribution} shows the cumulative histogram of FPs for all four combinations of codes and patterns. Pattern 40 clearly exhibits the highest FP rate in both {\sc mepsa} and {\sc fast-mepsa}. Having said that, the SNR distribution of FPs spans the range $3 \lesssim \mathrm{SNR} \lesssim 5$, indicating that all FPs lie in the region where the statistical significance of the pulse is uncertain. A possible way to address this issue is to impose a threshold on the SNR of detected pulses, requiring $\mathrm{SNR} > 5$, as it was done in our past investigations already mentioned above.

\subsection{True positive rate}
\label{subsec:tp}

Starting from Group 3 LCs, we computed the true positive (TP) rate for the four algorithm configurations, adopting two different SNR thresholds:
\begin{itemize}
    \item Case 1: $\mathrm{SNR} \ge 5$, to ensure the statistical significance of the detected peaks;
    \item Case 2: $4 \le \mathrm{SNR} < 5$, to assess the capability of the 40th pattern to recover peaks falling within the sub-threshold regime. 
\end{itemize}

\begin{table}[H]
\setlength{\tabcolsep}{8pt}
\renewcommand{\arraystretch}{1.4}
    \centering
    \begin{tabular}{c|cc}\hline
        Number of patterns & {\sc mepsa} & {\sc fast-mepsa} \\\hline
        39 & 29319 (276) & 28276 (160) \\
        40 & 34134 (1294) & 33225 (951) \\\hline
    \end{tabular}
    \caption{Number of TPs with $\mathrm{SNR} \ge 5$ ($4 \le \mathrm{SNR} < 5$) detected by each version of the algorithm in Group 3 LCs out of 76970 (8104) simulated peaks.}
    \label{tab:tp}
\end{table}

\begin{figure}[H]
    \centering
    \includegraphics[width=0.4\textwidth]{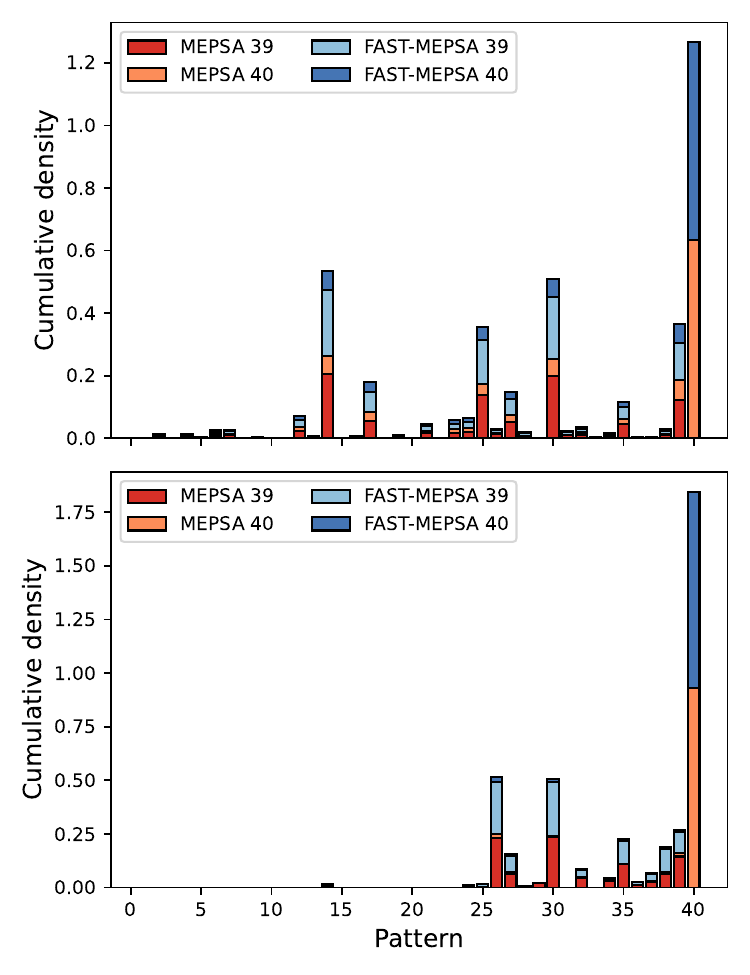}
    \caption{Top panel: Cumulative histogram of TPs with $\mathrm{SNR} \ge 5$ detected by all four algorithm configurations, as a function of the triggering pattern. Bottom panel: Cumulative histogram of TPs with $4 \le \mathrm{SNR} < 5$. In both panels, different colours represent the contribution of each algorithm to the total frequency in each histogram bin.}
    \label{fig:tp}
\end{figure}

The results, reported in Table~\ref{tab:tp}, show that {\sc fast-mepsa} leads to a slight reduction in the detection rate: approximately 4\% when using 39 patterns, and 3\% with 40 patterns. As done for FPs, we also examined the distribution of TPs across patterns. Figure~\ref{fig:tp} shows the TP spectra: the top panel displays the TP distribution for Case 1, while the bottom panel refers to Case 2. Interestingly, Figure~\ref{fig:tp} highlights the efficiency of the 40th pattern in detecting peaks in both regimes---particularly in the sub-threshold case, where it recovers a significantly larger fraction of pulses compared to the other patterns.

\subsubsection{Efficiency}
\label{subsubsec:efficiency}
The efficiency of each algorithm was computed using Group 3 LCs. \citetalias{Guidorzi15a} found that the relevant parameter to detect a given pulse $i$ is its separability $s_i$, defined as:
\begin{equation}
    s_i = \dfrac{\Delta t_{\min,i}}{\mathrm{FWHM}_i},
\end{equation}
where $\Delta t_{\min,i}$ and $\mathrm{FWHM}_i$ are the pulse’s minimum adjacent waiting time and its FWHM, respectively. For a pulse $i$ peaking at time $t_{\mathrm{p},i}$, the minimum waiting time is defined as $\Delta t_{\min,i} = \min(t_{\mathrm{p},i} - t_{\mathrm{p},i-1}, t_{\mathrm{p},i+1} - t_{\mathrm{p},i})$. Overall, the separability ranges from $-3 \lesssim \log{(s)} \lesssim 2$. Clearly, the smaller the separability between adjacent pulses, the harder it is for the algorithms to disentangle them, regardless of their SNR.

Following \citetalias{Guidorzi15a}, we divided the SNR--$s$ plane into 30 bins. In each bin, we computed the efficiency as the fraction of identified peaks over the total number of pulses. Figure~\ref{fig:efficiency} compares the four algorithm configurations. The recovered behaviour confirms the findings of \citetalias{Guidorzi15a}:
\begin{itemize}
    \item Efficiency is very high in the top-right region, where pulses are well separated and intense;
    \item At a given separability, efficiency increases slightly with SNR;
    \item When $\log{(s)} \lesssim -0.4$, pulses are hardly detected as separate structures, regardless of SNR;
    \item When $\log{(\mathrm{SNR})} \lesssim 0.7$, efficiency also drops significantly, nearly independent of separability.
\end{itemize}
The comparison shown in Figure~\ref{fig:efficiency} can be interpreted along both rows and columns:
\begin{itemize}
    \item Row-wise comparison: the efficiency in the SNR--$s$ plane is shown for the same algorithm version, either the original {\sc mepsa} or {\sc fast-mepsa}, before and after including the 40th pattern. In both rows, the inclusion of the new pattern visibly enhances the efficiency, particularly in the top right corner and, more generally, in the region within the range $\log\left(\mathrm{SNR}\right) \lesssim 1.5$ and $\log\left(s\right) \lesssim -1$. This behaviour is expected, since the 40th pattern was specifically designed to recover faint and elusive peaks.
    \item Column-wise comparison: we compare {\sc mepsa} and {\sc fast-mepsa} using the same number of patterns, 39 and 40, respectively. In both cases, the efficiency remains nearly identical, confirming that the modifications introduced in Section~\ref{subsec:mepsa_fast} do not affect the performance.
\end{itemize}

\begin{figure*}
    \centering
    \includegraphics[width=\textwidth]{figs/efficiency.pdf}
    \caption{Peak detection efficiency in the SNR--separability plane for {\sc mepsa} using 39 patterns (top left), {\sc mepsa} using 40 patterns (top right), {\sc fast-mepsa} using 39 patterns (bottom left), and {\sc fast-mepsa} using 40 patterns (bottom right). Different contour levels (from cold to hot colours) correspond to 20 different, equally spaced efficiency levels from 0 to 1.}
    \label{fig:efficiency}
\end{figure*}

\subsubsection{Visually identified missing peaks}
\label{subsubsec:missing_peaks}

We applied {\sc mepsa} with pattern 40 to the subset of visually identified peaks from \citet{Maccary24} that had been missed by the original pattern set. The inclusion of the new pattern allowed us to recover approximately 30\% of these primarily sub-threshold, low-SNR structures.

\begin{figure}[H]
    \centering
    \includegraphics[width=0.45\textwidth]{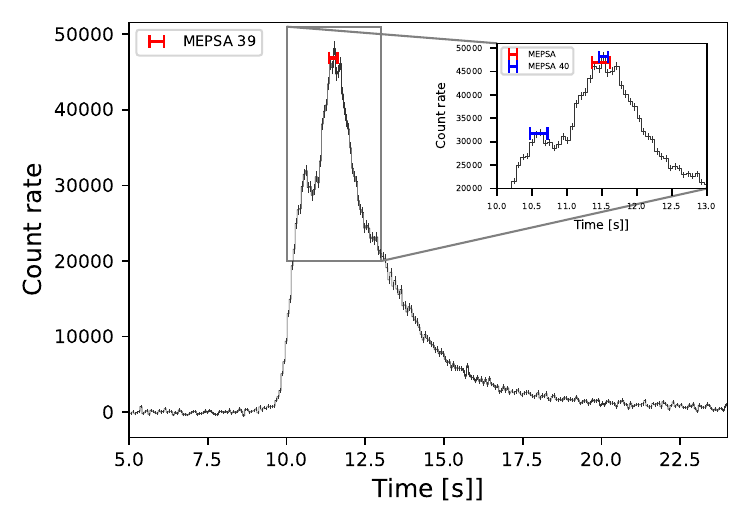}
    \caption{Example of a visually identified peak in GRB\,180728A that is missed by {\sc mepsa} with 39 patterns and successfully recovered with the inclusion of the 40th pattern. The zoomed-in time interval highlights the peak lying on the rising edge of a broader structure.}
    \label{fig:missing_peaks}
\end{figure}

Figure~\ref{fig:missing_peaks} shows an illustrative example from GRB\,180728A. The figure displays the output of {\sc mepsa} using 39 patterns, where one peak is clearly missed. The zoomed-in panel shows the same LC analysed with the 40th pattern included: the previously undetected structure is now successfully identified. This example demonstrates the effectiveness of the 40th pattern in capturing elusive peaks that lie on the rising edge of broader emission features.

\section{Discussion and conclusions}
\label{sec:conclusions}

Originally introduced by \citetalias{Guidorzi15a}, {\sc mepsa} was developed to identify peaks in GRB LCs, although it can be readily applied to other astrophysical transient phenomena. Specifically, {\sc mepsa} applies a set of predefined patterns simultaneously to an evenly sampled, background-subtracted input LC affected by uncorrelated Gaussian noise. The same set of patterns is then applied to re-binned versions of the input LC up to a maximum re-binning factor $F_{\rm reb,m}$. For each $F_{\rm reb}$, all possible temporal offsets are scanned, resulting in a computational time that scales approximately as $F_{\rm reb,m}^3$. This ultimately limits the maximum re-binning factor that can be practically used.

To overcome this limitation, we developed {\sc fast-mepsa}, an optimised version of the original algorithm that preserves its operative structure and philosophy. By adopting a sparser scanning strategy at high re-binning factors, {\sc fast-mepsa} achieves a reduction in computational time of nearly a factor of 400 compared to {\sc mepsa}. In parallel, we introduced a new 40$^{\rm th}$ pattern, specifically designed to recover a class of elusive and mostly sub-threshold peaks that preferentially lie on the rising edges of broader structures and are often missed by the original set.

We validated both combinations of codes ({\sc mepsa} and {\sc fast-mepsa}) with both sets of patterns (39 and 40) on the same set of simulated LCs that had originally been utilised for calibrating {\sc mepsa}.
Our results show that {\sc fast-mepsa} maintains almost the same detection efficiency as {\sc mepsa}, while significantly reducing the false positive (FP) rate thanks to its coarser offset scanning. This comes at the cost of a minor ($\sim 4\%$) reduction in the number of detected peaks. In addition, the introduction of the 40th pattern increased the detection rate, especially for low-SNR, sub-threshold events, demonstrating its usefulness in scenarios where faint signals may carry scientific interest, such as in multi-messenger follow-up searches.

Summing up, we propose a practical set of rules of thumb that may help the user to choose the optimal combination of code and set of patterns that best suites their needs.
\begin{itemize}
\item For large-scale analyses over wide datasets---where computational efficiency is crucial---{\sc fast-mepsa} is the best if not the only viable option. An example of application of {\sc fast-mepsa} is offered by its extensive use within a series of simulations based on a genetic algorithm (Maccary et al., submitted). The same recommendation also applies to the cases in which a wide range of timescales is to be scanned, which requires a large value of the maximum rebinning factor.
\item When the completeness of the sample of detected peaks is more important than purity, the 40-pattern set should be used instead of the 39-pattern one. This applies independently of the kind of code, either {\sc mepsa} or {\sc fast-mepsa}, with the caveat of the previous bullet point (for relatively short LCs and/or relatively small maximum rebin factors, {\sc mepsa} is to be preferred over {\sc fast-mepsa}). This is especially the case when the identification of sub-threshold peaks may be relevant.
\item When the purity of the sample is to be preferred over its completeness, we recommend to avoid using pattern 40. Concerning the code, there is a slight preference for {\sc fast-mepsa}, given its somewhat lower FP rate.
\end{itemize}

\section*{Acknowledgements}
M.M.~and R.M.~acknowledge the University of Ferrara for the financial support of their PhD scholarships.
%\appendix

\bibliographystyle{elsarticle-harv} 
\bibliography{alles_grbs}

\end{document}